\documentclass[apjl]{emulateapj}

\shortauthors{Chandar, Fall, \& Whitmore}
\shorttitle{Age Distributions of Clusters and Stars}

\def\lea{\mathrel{<\kern-1.0em\lower0.9ex\hbox{$\sim$}}}
\def\gea{\mathrel{>\kern-1.0em\lower0.9ex\hbox{$\sim$}}}
\newcommand{\lta}{{\>\rlap{\raise2pt\hbox{$<$}}\lower3pt\hbox{$\sim$}\>}}
\newcommand{\gta}{{\>\rlap{\raise2pt\hbox{$>$}}\lower3pt\hbox{$\sim$}\>}}

\begin{document}

\title{CONNECTION BETWEEN THE AGE DISTRIBUTIONS OF STAR CLUSTERS AND FIELD STARS: A FIRST APPLICATION TO THE SMALL MAGELLANIC CLOUD}

\author{Rupali Chandar$^{1,2}$, S. Michael Fall$^{3}$, and Bradley C. Whitmore$^{3}$}
\affil{$^1$ Center for Astrophysical Sciences, Johns Hopkins University, 3400 N. Charles Street, Baltimore, MD 21218 \\
\noindent $^2$ Carnegie Observatories, 813 Santa Barbara St., Pasadena, CA 91101-1292 \\
\noindent $^3$ Space Telescope Science Institute, 3700 San Martin Drive, 
Baltimore, MD 21218}

\begin{abstract}

We present the age distributions for star clusters and individual
stars in the Small Magellanic Cloud (SMC) based on data from the
Magellanic Clouds Photometric Survey by Zaritsky and collaborators.
The age distribution of the SMC clusters shows a steep decline,
$dN_{cluster}/d\tau \propto \tau^{-0.85\pm0.15}$, over the period $10^7
\lea \tau \lea 10^9$~yr.  This decline is essentially
identical to that observed previously for more massive clusters in the
merging Antennae galaxies, and also for lower-mass embedded clusters
in the solar neighborhood.  The SMC cluster age distribution therefore
provides additional evidence for the rapid disruption of star clusters
(``infant mortality'').  These disrupted clusters deliver their stars
to the general field population, implying that the field star age
distribution, $dN_{fld star}/d\tau$, should have an inverse relation
to $dN_{cluster}/d\tau$ if most stars form initially in clusters.  We
make specific predictions for $dN_{fldstar}/d\tau$ based on our
cluster disruption models, and compare them with current data
available for stars in the SMC.  While these data do not extend to
sufficiently young ages for a definitive test, they are consistent
with a scenario wherein most SMC stars formed in clusters.
Future analyses of $dN_{fldstar}/d\tau$ that extend down to ages of
$\sim$ few million years are needed to verify the age
relationship between stars residing in clusters and in the field.

\end{abstract}


\section{INTRODUCTION}

The age distribution of a population of star clusters contains
important information about the formation and disruption of the
clusters\footnote{We use the term ``cluster'' generically to mean any
aggregate of stars -- regardless of mass, size, or age, and whether
bound or unbound -- with a density significantly higher than the local
stellar background, making it recognizable as a distinct entity.}.
The number of embedded clusters in the solar neighborhood (with masses
$\sim10^2-10^3~M_{\odot}$) declines rapidly with age, and it has been
estimated that only $4-7$\% of these clusters will survive for 100~Myr
(see the review by Lada \& Lada 2003).  We recently found that much
more massive (compact) clusters ($\geq3\times10^4~M_{\odot}$) in the
merging ``Antennae'' galaxies also show a rapid decline in their age
distribution, with $dN_{cluster}/d\tau\propto\tau^{-1}$ (Fall,
Chandar, \& Whitmore 2005, hereafter FCW05).  We interpret this
decline in numbers as due to a high rate of early disruption (``infant
mortality'') for massive clusters in the Antennae.  Taken together,
these age distributions suggest that compact, massive clusters in the
chaotic environment of two merging disk galaxies disrupt at
approximately the same rate as lower mass, embedded clusters in the
more quiescent solar neighborhood.  This suggests that star clusters
disrupt rapidly due to processes internal to the clusters themselves
(e.g., the removal of interstellar material due to the mass and energy
input from massive stars; Hills 1980; Fall 2004; FCW05), and implies
that the age distribution should be roughly $dN_{cluster}/d\tau
\propto \tau^{-1}$ in all star-forming galaxies.

In contrast to the results for clusters in the Milky Way and in the
Antennae, the cluster age distribution in the Small Magellanic Cloud
(SMC) has been reported as $dN_{cluster}/d\tau\propto\tau^{-2.1}$
(Rafelski \& Zaritsky 2005; hereafter RZ05).  However, this age
distribution is actually the number of clusters divided by the number
of individual stars of the same age, i.e., it is a {\it normalized}
cluster age distribution (RZ05).  Does the proposed $\tau^{-2}$ power
law for the normalized cluster age distribution in the SMC differ from
the $\tau^{-1}$ form found in the Antennae and solar neighborhood
because of the normalization procedure, because clusters in the SMC
are forming and/or dissolving at a different rate, or for some other
reason?  The motivation for normalizing the number of clusters by the
number of individual stars was to assess whether the formation of
stars in clusters parallels the formation of stars in the field.  An
underlying assumption made by RZ05, then, is that star formation
occurs in {\it both} clusters and in the field simultaneously.
However, our study of the Antennae galaxies suggests that at least
20\% and possibly all of the $H\alpha$ flux is emitted by compact
clusters, which disrupt rapidly (FCW05).  This picture implies instead
that at least 20\% and possibly all stars form in clusters, with stars
being subsequently delivered from clusters to the field through the
process of cluster disruption.  This in turn implies that
$dN_{cluster}/d\tau$ and $dN_{fldstar}/d\tau$ should have an inverse
relationship to each other.

If the SMC cluster system is found to have an unnormalized age
distribution $\propto \tau^{-2.1}$, then our picture of clusters
disrupting at essentially the same rate in all star-forming
environments may have to be significantly revised.  What is the
unnormalized SMC cluster age distribution, does it differ from the
normalized cluster age distribution, and what is the relationship
between field and cluster stars?  Addressing these questions is the
main focus of this Letter.

\section{AGE DISTRIBUTION OF CLUSTERS IN THE SMC}

We use the SMC cluster properties published in RZ05 to directly determine
the cluster age distribution.  The original cluster sample
was selected by Hill \& Zaritsky (2006), and we refer the reader to
their paper for details.  Briefly, they identified cluster candidates
using $UBVI$ data from the Magellanic Clouds Photometric Survey (MCPS;
Zaritsky, Harris, \& Thompson 1997).  All candidate clusters detected
in median-subtracted stellar images were visually inspected, and only
obvious clusters were retained.  Clusters embedded in strong nebular
emission regions were excluded from the sample due to the difficulty
of measuring their true integrated colors.  This leads
to a significant bias against clusters younger than $\sim10^7$~yr.
Due to these selection criteria and the small distance to the SMC, the
cluster sample of Hill \& Zaritsky (2006) is likely to be
approximately surface-brightness limited, rather than
luminosity-limited (as is typical for more distant cluster systems).
The adopted procedure resulted in a catalog of 204 unambiguous SMC
clusters.

Basic cluster properties were estimated as follows.  RZ05 measured
integrated $UBVI$ magnitudes within the radius enclosing 90\% of the
V-band light from each cluster (as tabulated in Hill \& Zaritsky
2006).  They then compared the colors (typically $U-B$, $B-V$, and
$V-I$) with the simple stellar population (SSP) model predictions of
STARBURST99 (Leitherer et al. 1999) and GALEV (Anders \&
Fritze-v. Alvensleben 2003) and adopted the age corresponding to the
minimum $\chi^2$.  RZ05 provide electronic tables of derived cluster
ages and total V band luminosities for the clusters in their catalog.
We estimated the mass of each cluster by combining the total V band
luminosity provided by RZ05 with theoretical, age-dependent SSP
mass-to-light ratios ($M/L_V$), assuming a distance modulus of 18.87
(Harries et al. 2003) and a foreground reddening value of
$E_{B-V}=0.09$.  These $M/L_V$ ratios assume a Salpeter initial
stellar mass function (IMF) down to $0.1~M_{\odot}$, but would be
$\sim1.5$ times lower if we had adopted a Chabrier-type (2003) IMF
that flattens below $1~M_{\odot}$.  Our procedure does not account for
any additional extinction within the SMC, which is likely higher for
younger clusters, resulting in an underestimate for the masses of
these objects.  However, the extinction in the SMC is quite small
along most lines of sight (e.g., Zaritsky 1999), and adopting a single
extinction value has little effect on our results.

From the above data we constructed the age distribution of star
clusters in the SMC, $dN_{cluster}/d\tau$, by counting the number of
clusters ($N$) in different age bins.  In this Letter, we use cluster
ages taken from the RZ05 comparison with the STARBURST99 model
(Leitherer et al. 1999) of metallicity Z=0.004, which is well-suited
for the young cluster population in the SMC.  However, we have checked
that our basic results are not sensitive to the adopted metallicity,
population synthesis model, or bin size.  The $1\sigma$ uncertainties
are calculated as $\pm\sqrt{N}$ for each bin.  The data were
restricted to cover the age range over which the sample of clusters is
reasonably complete down to $\sim10^3~M_{\odot}$.

\begin{figure}
\plotone{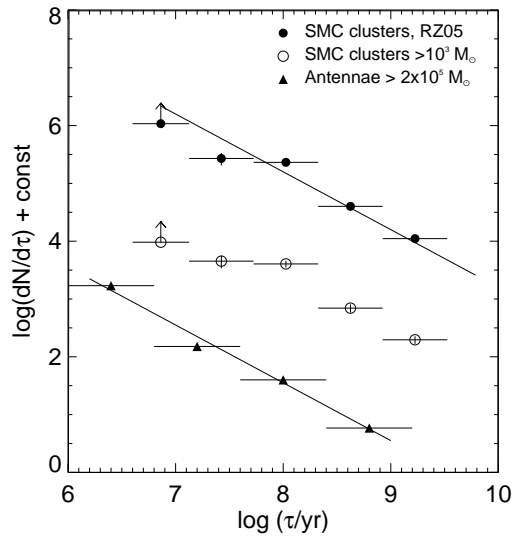}
\caption{
Age distributions of star clusters in the SMC and the Antennae.  
The different symbols refer to 
the approximately surface-brightness limited SMC cluster sample of RZ05
(filled circles), a mass-limited subset of the RZ05 cluster sample
(open circles), and massive star clusters in the Antennae (filled
triangles) as presented in FCW05.  The data point for the
youngest bin in the SMC cluster age distributions is a lower limit, due
to the cluster selection procedure (see text).  The solid lines
represent the power law $dN_{cluster}/d\tau \propto \tau^{-1}$. }
\end{figure}

Figure~1 shows the age distributions resulting from the full RZ05
sample with $\mbox{log} (\tau/\mbox{yr}) \lea 9.5$, and from the
approximately mass-limited sample ($\geq 10^3~M_{\odot}$).  As
expected, the age distribution constructed from the mass-limited
sample is somewhat shallower than the one constructed from the entire
cluster sample, since there are relatively few clusters more massive
than $10^3~M_{\odot}$ at very young ages.  Both samples have a median
age $\sim10^8$ yr.  This is a generous upper limit on the true median
age due to the original selection, which deliberately excluded
clusters in emission-line regions, with ages $\tau \lea 10^7$~yr.
Regardless of this incompleteness at very young ages, the age
distributions of Figure~1 show that in the SMC the number of clusters
declines monotonically, with no obvious bends or other features.  {\it
We find that the (unnormalized) cluster age distribution in the SMC is
well approximated by a power law of the form $dN_{cluster}/d\tau
\propto \tau^{-0.85\pm0.15}$, over the range of ages studied [$10^7 \lea
\tau \lea 10^9$~yr].}  This form of the SMC cluster
age distribution is essentially the same as that found for star
clusters in the Antennae galaxies (FCW05).  Our result therefore,
provides additional evidence for the rapid disruption of clusters due
to internal processes (``infant mortality'').

\section{FIELD STAR AGE DISTRIBUTION AND RELATIONSHIP TO THE CLUSTERS}

This rapid decline of $dN_{cluster}/d\tau$ with $\tau$ naturally
predicts that $dN_{fldstar}/d\tau$ should increase with $\tau$ as
dissolving clusters contribute their stars to the field.  However the
exact form of $dN_{fldstar}/d\tau$ will depend on the fraction of
stars that originally form in clusters.  In order to explore the
relationship between stars in clusters and in the field, we use the
statistical cluster models presented in Whitmore, Chandar, \& Fall
(2006; hereafter WCF06) to predict the field star age distribution.
Here, we provide a brief summary of the model inputs and assumptions,
and refer the reader to WCF06 for details.  The models allow different
cluster formation histories, including a constant rate of formation
and gaussian bursts.  For simplicity, we assume a constant rate of
cluster formation over the life of the SMC, and an initial mass
function of power-law form with $dN_{cluster}/dM \propto M^{-2}$ as
suggested by a number of recent works (e.g., Zhang \& Fall 1999;
Hunter et al. 2003; WCF06).  After clusters form, they are disrupted
in two phases -- a rapid, mass-independent phase (which is modelled as
a power law $dN/d\tau \propto \tau^{\gamma}$, with index $\gamma$),
and a longer-term, mass-dependent phase that mimics the effects of
two-body evaporation (modelled as constant mass loss, with a rate
$\mu_{ev}=1\times10^{-5}~M_{\odot}~\mbox{yr}^{-1}$; see Fall \& Zhang
2001 and WCF06 for details).  The index of the SMC cluster age
distribution was found in the previous section to be
$\gamma=-0.85\pm0.15$.  Each simulated cluster is composed of a number
of stars, which are assumed to be drawn from a Salpeter IMF with lower
and upper stellar mass limits of $0.1~M_{\odot}$ and $100~M_{\odot}$
respectively.  The only free parameter in our model is the fraction of
stars, $f$, that are born initially in clusters\footnote{Our idealized
calculations divide stars artificially into two distinct categories,
i.e., ``cluster'' and ``field''.  In reality, stars likely form in a
continuum of environments from tightly to more loosely clustered
aggregates.}.  The models track the numbers and ages of stars in the
field (i.e., stars which form outside of clusters initially or end up
there after a cluster is disrupted), thus allowing us to predict
$dN_{fldstar}/d\tau$.

Figure~2 shows our model predictions of $dN_{fldstar}/d\tau$ for three
different values of $f$: 0.0, 0.5, and 1.0.  The (solid) curve for
$f=1.0$ (i.e., {\it all} star formation occurs initially in clusters) shows a
steep rise in the number of field stars (at very young ages), which is
approximately the inverse of $dN_{cluster}/d\tau$.  However, by
$\sim10^7$~yrs the field star age distribution is essentially flat
(i.e., $dN_{fldstar}/d\tau \propto \tau^{0}$).  This effect can be
understood as follows: when $dN_{cluster}/d\tau$ is plotted
logarithmically, the {\it rate} at which the cluster population
diminishes is observed as the slope, $\gamma$.  However, if
$dN_{cluster}/d\tau \propto \tau^{-0.85}$ were plotted linearly rather
than logarithmically, one would see that the number of clusters
declines quickly at first, and then begins to flatten out.  The field
star age distribution is roughly the inverse of the cluster age
distribution, and so must necessarily rise very quickly as stars are
dispersed into the field.  This increasing function essentially
``saturates'' by $\sim10^7$~yrs, predicting that $dN_{fldstar}/d\tau$
varies rapidly only for a short time.

\begin{figure}
\plotone{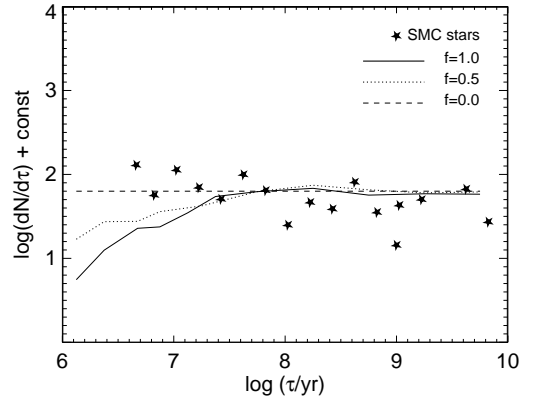}
\caption{Comparison between the SMC field star age distribution
(stars) from HZ04 and predictions for $dN_{fldstar}/d\tau$ from the
WCF06 models (see text for details).  The power-law index $\gamma$,
which quantifies the rate of cluster disruption, is fixed at $-0.85$,
as measured from the SMC cluster age distribution.  The only free
parameter in the model is the fraction of stars, $f$, 
that form within clusters.  Note that even if
100\% of stars form initially in clusters, the field star distribution
flattens by $\sim10^7$~yrs.  }
\end{figure}

Is the field star age distribution in the SMC consistent with most stars
forming in clusters?  Harris \& Zaritsky (2004; hereafter HZ04)
present the star formation history of the SMC.  They use all detected
stars (i.e., in both clusters and in the field), from the MCPS
ground-based data.  However, there is a natural bias in their stellar
catalog against stars in clusters, due to their inability to
photometer in dense regions.  Therefore, we refer to stars in their
sample as belonging to the ``field''.  We determine the number of SMC
stars as a function of age by dividing the star formation rates
presented in Table~2 of HZ04 by the mean stellar mass.  Figure~2 shows
a comparison of the field star age distribution in the SMC with our model
predictions.  This shows that over the observed range the field star
age distribution is essentially flat (although see HZ04 for a detailed
analysis of small-scale variations in the star formation history).
Unfortunately, the youngest age in the HZ04 compilation is just where
the predicted rise in the field star age distribution begins to
flatten, making it impossible to determine $f$ with this data set.
Additionally, the HZ04 data do not discriminate between stars in
clusters and stars in the field, which results in ``pollution'' of
$dN_{fldstar}/d\tau$ from cluster stars.  Nevertheless, Figure~2 shows
that despite $dN_{fldstar}/d\tau$ being rather flat, {\it the SMC
field star data do not contradict our scenario where most star
formation occurs in short-lived clusters}.  Our simulations
establish that even if 100\% of stars formed originally in clusters,
their rapid rate of disruption creates the field star population so
quickly that after $\sim10^7$~yrs the two populations will always
appear to evolve independently.

\section{DISCUSSION AND CONCLUSIONS}

We have shown here that the (unnormalized) cluster age distribution in
the SMC has the form $dN_{cluster}/d\tau \propto \tau^{-0.85\pm0.15}$
and that overall the field star age distribution is nearly flat (i.e.,
$dN_{fldstar}/d\tau\propto \tau^{0}$, over the age range $10^7 \lea
\tau \lea 10^9$~yr).  The {\it normalized} SMC
cluster age distribution (the number of clusters divided by the number
of individual stars of the same age) therefore must have the same form
as the unnormalized distribution [i.e., $(dN_{cluster}/d\tau)_{norm}
\propto \tau^{-0.85\pm0.15}$].  Our conclusion contradicts that of
RZ05, who found that the normalized cluster age distribution is
$(dN_{cluster}/d\tau)_{norm} \propto \tau^{-2.1}$ from the same data.
The difference in the power-law exponent between our work and that of
RZ05 comes about because we divided the cluster age distribution
($dN_{cluster}/d\tau$) directly by that for the field stars
($dN_{fldstar}/d\tau$), whereas RZ05 divided
$dN_{cluster}/d\mbox{log}\tau$ by $dN_{fldstar}/d\mbox{log}\tau$, and
then appear to have introduced an extraneous factor of $\tau^{-1}$.

Our result that the age distribution of SMC clusters has a power-law
form $dN_{cluster}/d\tau \propto \tau^{-0.85\pm0.15}$ has important
implications regarding the physical processes that disrupt star
clusters.  The fact that clusters in vastly different environments
(including the chaotic merging of two galaxies, a quiescent spiral
disk, and a gas-rich dwarf galaxy) have age distributions with nearly
identical forms provides strong evidence that star clusters disrupt
rapidly due to processes {\it internal} to the clusters themselves.
Over the first few million years photoionization, stellar winds, jets
and supernovae can remove much of the interstellar matter from
clusters leaving the member stars freely expanding (e.g., Hills
1980; Boily \& Kroupa 2003a,b).  These expanding clusters will be
observable for $\sim$~few~$\times 10^{7}$~yrs after becoming unbound,
until their surface brightness becomes so low that they become
indistinguishable from statistical fluctuations of stars in the field
(FCW05).  In addition to the removal of interstellar material, mass
loss from stars themselves will likely contribute to the continued
unbinding of clusters (e.g., Applegate 1986; Chernoff \& Weinberg
1990; Fukushige \& Heggie 1995).  Finally, two-body evaporation will
destroy any remaining clusters on longer timescales (e.g., Vesperini
1997; Baumgardt 1998; Fall \& Zhang 2001).  The early disruption
(``infant mortality'') appears to operate relatively independent of
cluster mass, while two-body evaporation is mass-dependent.

The stars from disrupted clusters add to the general field, requiring
a close relationship between the age distributions of clusters and
field stars if most stars form initially in clusters.  In this work we
compared age distributions for clusters and individual stars in the
SMC with model predictions to further test this general framework.
Our models suggest that the rapid drop in the cluster age distribution
should be reflected in a rapid rise in the field star age
distribution.  However this rapid rise in the number of field stars is
only observable for the first $\sim10^7$~yrs {\it regardless of the
fraction of stars that initially formed in clusters}.  Because the
present data on SMC stars begin just as the predicted field star
distributions flatten, and because there is no discrimination between
stars belonging to clusters versus to the field, they do not provide
an unambiguous answer.  However the observed relations are consistent
with a scenario where most stars form in clusters that are
subsequently disrupted by the mechanisms discussed above.  The models
also predict that even if all stars form initially in clusters,
galaxies will contain a relatively small number of observable clusters
at any given time.  Future studies with the {\it Hubble Space
Telescope} (which allow better discrimination between stars in
clusters and in the general field), and in the infrared (which can
probe very young, deeply embedded stars and clusters) could be used to
search for the expected rise in the field star age distribution at
$\tau \lea 10^7$~yrs.  A measurement of the rise in
$dN_{fldstar}/d\tau$ would provide a direct estimate of the fraction
of stars which originally form in clusters versus in the field.

\acknowledgments

We thank Jason Harris for explaining the HZ04 table of SMC star
formation rates, Dennis Zaritsky for discussions about the SMC cluster
age distribution, and Mike Santos and Francois Schweizer for helpful
comments on the manuscript.  R. C. is grateful for support from NASA
through grant GO-09192-01-A from STScI, which is operated by AURA,
Inc., under NASA contract NAS5-26555.

\end{document}